\documentclass[onecollarge]{svjour2}       % onecolumn "king-size"
\smartqed  % flush right qed marks, e.g. at end of proof
\usepackage{graphicx}
\usepackage{amssymb,amsmath}
\usepackage{graphicx}
\usepackage{bm}
\usepackage{color}

\begin{document}

\title{On the merit of a Central Limit Theorem-based approximation in statistical physics}
%\subtitle{Do you have a subtitle?\\ If so, write it here}

%\titlerunning{Short form of title}        % if too long for running head

\author{B. Leggio        \and
        O. Lychkovskiy   \and
        A. Messina
}

%\authorrunning{Short form of author list} % if too long for running head

\institute{B. Leggio \at
              Dipartimento di Fisica, Universit\`{a} degli Studi di Palermo, Via Archirafi 36, 90123 Palermo,
Italy \\
              \email{bruno.leggio@unipa.it}           %  \\
%             \emph{Present address:} of F. Author  %  if needed
           \and
           O. Lychkovskiy \at
              Institute for Theoretical and Experimental Physics, 117218, B. Cheremushkinskaya 25, Moscow, Russia
              \and
              A. Messina \at
              Dipartimento di Fisica, Universit\`{a} degli Studi di Palermo, Via Archirafi 36, 90123 Palermo,
Italy
}

\date{Received: date / Accepted: date}
% The correct dates will be entered by the editor

\maketitle

\begin{abstract}
The applicability conditions of a recently reported Central Limit Theorem-based approximation method in statistical physics are investigated and rigorously determined. The failure of this method at low and intermediate temperature is proved as well as its inadequacy to disclose quantum criticalities at fixed temperatures. Its high temperature predictions are in addition shown to coincide with those stemming from straightforward appropriate expansions up to $(k_B T)^{-2}$. Our results are clearly illustrated by comparing the exact and approximate temperature dependence of the free energy of some exemplary physical systems.
% \subclass{MSC code1 \and MSC code2 \and more}
\end{abstract}

\section{Introduction}
The thermodynamics of most everyday-scale phenomena, e.g phase transitions, often reflects the underlying existence of a complex microscopic many-body dynamics which is approachable, generally speaking, in statistical terms only. Understanding the collective behavior of physical systems comprising a myriad of interacting or not, classical or quantum particles does constitute an endless challenge, over the last years fed by the astonishing progresses in the production of new materials having properties tailored on demand \cite{mat3,mat2,mat1}, as well as in the manipulation of mesoscopic  quantum systems \cite{qpt,single}. Experimental and technological successes of this kind do indeed open new problems on a variety of many-body systems \cite{manybody,stamp}, to be dealt with by making recourse to the ideas and methods of statistical physics (see for example \cite{therm,fluct}). The correspondent task of the theoreticians attracted by these problems is to contrive simple enough but well addressed microscopic models as well as to develop, within the Gibbs framework, new widely applicable mathematical/numerical tools leading to predictions of experimental interest. Numerical approaches based on the Renormalization Group (RG) method \cite{rg,rg2} have been successfully employed to produce a deeper insight into the behavior of quantum matter. Such applications are even more important in the cases of non-integrable models when an exact (even if usually complicated) solution is not available. Nevertheless these methods often require a huge computational effort, and might thus be heavily computer-based \cite{rg3}.

Recently a different approach based on a generalized version of the Central Limit Theorem (CLT) has been proposed by Hartmann, Mahler and Hess \cite{lim} and, at least for some specific class of problems, it turns out to be much simpler in its application than the RG-based treatment. Such an alternative method relies on a gaussian approximation for the distribution of physical properties of quantum systems and in what follows we will refer to it as the Gaussian Approximation (GA). In this paper we critically review such a method, clarifying its limits and highlighting its range of applicability, also providing applications to some exemplary physical system. In particular we show that this method has good agreement with the exact thermodynamic behavior only in the limit of high temperature, while it fails in reproducing the physics of a wide class of systems at low and intermediate temperatures. The existence of limitations for this method has already been noticed in \cite{lim}, where the authors show that GA is unable to predict the true behavior of a particular system, namely an Ising chain, at low temperature, and suggest that this should be a generic feature of GA. The novelty of our work is that, for the first time, it is rigorously shown the existence of systematic drawbacks in the applicability of GA. In particular our analysis confirms, elucidates and generalizes the limitations of this approach as discussed in \cite{lim}.

This paper is organized as follows: Section 2 is devoted to a brief review of GA. Mathematical techniques allowing for closed expressions for thermodynamical quantities in the framework of GA are developed in Section 3. The method is applied in Section 4 to obtain results on the thermodynamics of spin models, paradigmatic examples of many-body quantum systems. Comments and conclusions are drawn in Section 5.

\section{Gaussian Approximation}
Here we briefly review the Gaussian method, originally presented in \cite{lim} and based on a mathematical theorem proven in \cite{ga}. The notations employed strictly follow \cite{lim}. For more mathematical details on the fundamental theorem underlying the approximation and its proof we refer the interested reader to \cite{ga} and to references therein.

Let us consider an $N$-block quantum linear chain with Hamiltonian $H=\sum_{\mu=1}^{N}\mathcal{H}_{\mu}$, where $\mathcal{H}_{\mu}$ is an operator composed of the block self-Hamiltonian and the interaction between blocks $\mu$ and $\mu+1$, thus reading
\begin{equation}\label{block hamiltonian}
\mathcal{H}_{\mu}=\mathbb{I}^{\otimes (\mu-1)}H_{\mu}\mathbb{I}^{\otimes (N-\mu)}+\mathbb{I}^{\otimes (\mu-1)}I_{\mu,\mu+1}\mathbb{I}^{\otimes (N-\mu-1)},
\end{equation}
$\mathbb{I}$ being the identity operator in the single-block Hilbert space.
Each block is described by a finite-dimensional Hilbert space with dimensionality $d$, the dimension of the total Hilbert space thus being $d^N$. It may consist e.g. of one or several neighboring spins. Given a total Hamiltonian $H$, one may define the partitioning into blocks in many different ways. It is worth stressing, however, that GA applies as long as the dimension of Hilbert space of each block stays finite.

Let us now define two basis of the total Hilbert space of the system. The first one, whose vectors will be written as $| \phi \rangle$, is composed by eigenstates of $H$ such that $H|\phi\rangle=E_{\phi}|\phi\rangle$. The second basis is made up of factorized vectors, written as $|a\rangle = \prod_{\mu}|a_{\mu}\rangle$, where each $|a_{\mu}\rangle$ is a local state of block $\mu$. Finally, we define the two Gaussian parameters
\begin{equation}\begin{split}
\overline{E_{a}}&=\langle a| H |a\rangle, \\
\Delta_{a}^{2}&=\langle a| H^{2}|a\rangle - \langle a| H |a\rangle^{2}.
\end{split}\end{equation}
To calculate free energy and other thermodynamic quantities in Gaussian approximation, one needs to know only these two parameters and the ground state energy. Let us for a moment leave apart the problem of calculating the ground state energy (which may be solved sometimes exactly and otherwise numerically). It is often an easy task to calculate the above two gaussian parameters, as will be demonstrated on some examples below. For this reason Gaussian method may seem apparently more attractive than other, much more sophisticated methods.

From the definitions given above, it is possible to build the operator
\begin{equation}
\zeta=\frac{H-\overline{E_{a}}}{\Delta_{a}}
\end{equation}
with eigenvalues $z_{\phi}$.
It was shown in \cite{ga} with the use of a Central Limit Theorem, that under the condition that a constant $C >0$ exists such that
\begin{equation}\label{conditions for QCLT}
\Delta_{a}^{2}\geq NC,
\end{equation}
the following limit strictly holds:
\begin{equation}\label{palim}
\lim_{N\rightarrow \infty}\mathrm{P}_{a}(z_{\phi}\in [z_{1},z_{2}])=\int_{z_{1}}^{z_{2}}\frac{e^{-\frac{z^{2}}{2}}}{\sqrt{2\pi}}dz,
\end{equation}
where $\mathrm{P}_{a}(z_{\phi}\in [z_{1},z_{2}])\equiv\sum_{\{ |\phi\rangle:z_{1}\leq z_{\phi}\leq z_{2} \}}|\langle a|\phi\rangle|^{2}$.\\

If one naively applies this result for calculation of density of states $\eta_{G} (E)$ and partition function $Z_{G}$, one readily obtains the following expressions \cite{lim}, which constitute what we call the Gaussian Approximation:
\begin{equation}\label{dens}
\begin{split}
\eta_{G}(E)&=\sum_{\{ |a\rangle \}}\frac{e^{\frac{(E-\overline{E_{a}})^{2}}{2\Delta_{a}^{2}}}}{\Delta_{a}\sqrt{2\pi}},
\end{split}
\end{equation}
\begin{equation}\label{part}
\begin{split}
Z_{G}&=\sum_{\{ |a\rangle \}}\frac{1}{2}e^{(-\beta \overline{E_{a}}+\frac{\beta^{2}\Delta_{a}^{2}}{2})}\Bigg( \mathrm{erfc}\Big( \frac{E_{g}-\overline{E_{a}}+\beta \Delta_{a}^{2}}{\Delta_{a}\sqrt{2}} \Big)\Bigg),
\end{split}
\end{equation}
where $E_g$ is the ground state energy of the system, the summation is extended over the whole factorized basis $\{ |a\rangle \}$ and $\mathrm{erfc}(x)$ is the conjugate gaussian error function defined as
\begin{equation}
\mathrm{erfc}(x)\equiv\frac{2}{\sqrt{\pi}} \int_x^\infty e^{-u^2}du.
\end{equation}
Eq. \eqref{part} is obtained from eq. \eqref{dens} via
\begin{equation}\label{Z from eta}
\begin{split}
Z_{G}&=\int_{E_g}^{+\infty} \eta_{G}(E) e^{-\beta E} dE.
\end{split}
\end{equation}
Here $\beta$ is the inverse temperature in units of $k_B$.\\
It should be emphasized that rigorously speaking eq. \eqref{palim} does not imply eqs. \eqref{dens}, \eqref{part}. The reason is that the convergence in \eqref{palim} is the convergence of a distribution, which is too weak to justify the summation of large (infinite, in the limiting case) number of terms \cite{lim}. Thus eqs. \eqref{dens}, \eqref{part} must be considered as an approximation motivated by a Central Limit Theorem, whose range of validity should be carefully established. This latter task is the goal of our work.

In performing such an analysis, we often will work in the limit of very high number of blocks, when the number of terms in the sum in eqs. \eqref{dens}, \eqref{part} becomes very large and goes to infinity in the thermodynamic limit. In these situations it is reasonable to exchange summation over $\{|a\rangle\}$ by integration over energy, introducing a density $\eta_{free}(\overline{E})$ defined as the number of states $|a\rangle$ for which $\overline{E_{a}}\in[\overline{E},\overline{E}+d\overline{E}]$. With this definition one can then replace summation with integration as
\begin{equation}\label{sumint}
\sum_{\{ |a\rangle \}} \rightarrow \int \eta_{free}(\overline{E}) d\overline{E}.
\end{equation}

\section{GA and exact physical predictions: a comparison}
In this Section we are interested in studying the physical predictions of GA and comparing them with the true behavior of thermodynamic quantities. In what follows we try to be as general as possible, in order to analyze the how and why GA shows limits of applicability. In the next section we will show how these limits apply to some particular system.

\subsection{Error function at large positive and negative arguments}
To begin with, we highlight a technical issue which is crucial for discussing GA at large $N$. Let us focus on the behavior of complementary error function which enters eq. (\ref{part}). Its argument grows with number of blocks as $\sqrt{N}$ \cite{hmh}. Indeed, one expects that
\begin{equation}\label{linera scaling}
\begin{array}{c}
E_g=N \varepsilon_g (1+O(N^{-1})),\\
\overline{E_a}=N \overline{\varepsilon}_a (1+O(N^{-1})),\\
\Delta_a^2=N \delta_a^2 (1+O(N^{-1})),\\
\end{array}
\end{equation}
where $\varepsilon_g,$ $\overline{\varepsilon}_a$ and $\delta_a$ are quantities defined {\it per block} and thus do not depend on $N.$ The above linear scaling is a generic feature of a modular Hamiltonian (\ref{block hamiltonian}). This will be exemplified by means of specific models in Sec. \ref{sec specific models}. Also note that linear scaling of  $\Delta_a^2$ is in accordance with the condition (\ref{conditions for QCLT}). So the argument of the complementary error function in eq. (\ref{part}) can be written as
\begin{equation}
\sqrt{N}\frac{\varepsilon_{g}-\overline{\varepsilon_{a}}+\beta \delta_{a}^{2}}{\delta_{a}\sqrt{2}}.
\end{equation}
At large N this is a large number, either positive or negative depending on the sign of $(\varepsilon_{g}-\overline{\varepsilon_{a}}+\beta \delta_{a}^{2})$ \cite{hmh}. Complementary error function has a well-known asymptotic behavior at large arguments:
\begin{equation}\label{erfc asymptotic}
{\rm erfc}(x)\sim
\left\{
\begin{array}{lcl}
2, & x\rightarrow -\infty,\\
 e^{-x^2}/(\sqrt{\pi}x), & x\rightarrow +\infty.\\
\end{array}
\right.
\end{equation}
As we will see this different behavior will produce certain discontinuities at the temperature where $(\varepsilon_{g}-\overline{\varepsilon_{a}}+\beta \delta_{a}^{2})$ changes its sign. In what follows we will need not only the above asymptotic relations, but also an exact inequality \cite{ineq} which limits ${\rm erfc}(x)$ at $x>0:$
\begin{equation}\label{erfc inequality}
{\rm erfc}(x) <  e^{-x^2}/(\sqrt{\pi}x).
\end{equation}

\subsection{High temperature limit}
To begin with, we want to compare the high temperature behaviors of both the exact partition function $Z=\mathrm{Tr}\big( e^{-\beta H} \big)$ and the partition function obtained from GA. Formally expanding the exponential function of $H$ in $Z$ up to the second order in $\beta$ one straightforwardly gets
\begin{equation}\label{exactZhighT}
Z= \mathrm{Tr}\Big( 1-\beta H + \frac{\beta^2}{2}H^2 \Big)+O(\beta^3).
\end{equation}
Note now that such a trace is independent on the basis chosen to practically evaluate it. Thus one can use the factorized basis $\{ |a\rangle \}$ defined in the previous section.
In this way, equation \eqref{exactZhighT} can be written as
\begin{equation}\label{exactZhighTfact}\begin{split}
Z&= \sum_{\{ |a\rangle \}}\Big( 1-\beta \langle a |H|a\rangle +\frac{\beta^2}{2}\langle a|H^2|a\rangle\Big)+O(\beta^3) \\
&=\sum_{\{ |a\rangle \}}\Big( 1-\beta \langle a |H|a\rangle +\frac{\beta^2}{2}\langle a|H|a\rangle^2 +\frac{\beta^2}{2}\big( \langle a | H^2|a\rangle - \langle a|H|a\rangle^2 \big) \Big)+O(\beta^3).
\end{split}\end{equation}
Equation \eqref{exactZhighTfact} does not of course coincide with \eqref{part}. Nevertheless, in the limit of high temperature,
\begin{equation}
T>\max_{\{ |a\rangle \}}
\frac{\delta_{a}^{2}}{\overline{\varepsilon_{a}}-\varepsilon_g}
,
\end{equation}
arguments of {\it all} error functions in \eqref{part} are positive, and one can use the asymptotic form of $\mathrm{erfc}(x)$ at $x \rightarrow -\infty,$ see eq. (\ref{erfc asymptotic}), to obtain
\begin{equation}\label{hightexp}\begin{split}
Z_G&\simeq \sum_{\{ |a\rangle \}}e^{\beta^2 \frac{\Delta_a^2}{2}}e^{-\beta \overline{E_{a}}}\\
&=\sum_{\{ |a\rangle \}}\Big( 1-\beta \langle a |H|a\rangle +\frac{\beta^2}{2}\langle a|H|a\rangle^2 +\frac{\beta^2}{2}\big( \langle a | H^2|a\rangle - \langle a|H|a\rangle^2 \big) \Big)+O(\beta^3),
\end{split}\end{equation}
which is the same result obtained in \eqref{exactZhighTfact}.

It is thus shown that GA and exact results coincide, up to the order $\beta^2$, in the high temperature limit, since by exploiting the knowledge of partition function one can obtain expressions for every thermodynamical quantity of interest.
\\ \\

\subsection{Low temperature limit}

What about low temperatures?
In order to avoid unwanted divergences when taking the limit $T\rightarrow 0$, in the present subsection (but nowhere else!) we define the Hamiltonian (\ref{block hamiltonian}) in such a way that the ground state energy $E_g$ equals zero.
Then the exact partition function reads
\begin{equation}\label{pf}
Z=1 + \sum_{i=2}^{d^N} e^{-\beta E_i} > 1.
\end{equation}
Here $E_i$ are eigenstates of the total Hamiltonian.

On the other hand, at low enough temperatures,
\begin{equation}
T<\min_{\{ |a\rangle \}}
\frac{\delta_{a}^{2}}{\overline{\varepsilon_{a}}}
,
\end{equation}
one is able to limit the approximate partition function $Z_G$ from above. Indeed at such low temperatures arguments of {\it all} error functions in eq. (\ref{part}) are negative, and  exploiting inequality (\ref{erfc inequality}) one obtains
\begin{equation}\label{pfg}
Z_{G} \leq d^N \max_{\{ |a\rangle \}}
\frac{e^{-\overline{\varepsilon}_a^2/(2\delta_a^2)}}{\sqrt{\pi N/2}(\beta\delta_a-\overline{\varepsilon}_a/\delta_a)}
\rightarrow 0 ~~{\rm as}~~ T\rightarrow0.
\end{equation}
We thus see how the approximate partition function can not reproduce the exact result when temperature is low enough. This is a consequence of the fact that the Gaussian Approximation does not hold when the system is very close to being in a single eigenstate of $H$ (in our case, in the ground state) \cite{lim}. In \cite{lim} such a failure in reproducing exact results at $T\approx 0$ was noted in connection with the study of a Ising chain, and it was argued that GA would in general fail when predicting physical properties at zero temperature.

Up to now we have conducted a very general analysis of GA at high and low temperatures, which has revealed that while at high temperatures this approximation works satisfactory, at low temperatures something goes wrong. In the next sections we show that under specific constraints on gaussian parameters $\overline{E_a}$ and $\Delta_a$ it is possible to give explicit expressions for measurable quantities obtained from $Z_G$ in the thermodynamic limit in the {\it whole range of temperatures}.
This analysis will allow us to elucidate that, contrarily to what happens in the high temperature limit, GA is unable to reproduce physical results in the range of low and intermediate temperatures.

\subsection{An analytical expression for the free energy under constant Gaussian parameters}
Consider a model in which Gaussian parameters $\Delta_a$ and $\overline{E_{a}}$ in some factorized basis do not depend on $a$:
\begin{equation}\label{assumptions}\begin{split}
\overline{E_{a}}&=0,\\
\Delta_{a}&=\sqrt{N}\delta>0~~\forall~a.
\end{split}\end{equation}
Note that the first equation implies the ground state energy to be negative unless the Hamiltonian is identically zero.
From equation \eqref{part} it is then possible to obtain the Gaussian Approximation for the density of states and the partition function:
\begin{equation}
\label{densconst}
\begin{split}
\eta_{G}& =\frac{d^N}{\sqrt{2\pi N}\delta} e^{-E^2/(2\delta^2 N)},\\
Z_{G}&=\frac{d^N}{2}e^{\frac{\beta^{2}\delta^{2}N}{2}}\mathrm{erfc}\bigg[\sqrt{\frac{N}{2}}\Big(\frac{\delta}{T}+\frac{\varepsilon_g}{\delta}\Big)\bigg].
\end{split}
\end{equation}

We wish to calculate free energy per particle which is defined as $-N^{-1}T \ln Z.$ In the GA framework it can be easily obtained from eq. \eqref{densconst} with the use of asymptotic expressions \eqref{erfc asymptotic} for error function. In the thermodynamic limit the result reads
\begin{equation}\label{fG}
f_G\equiv \lim_{N\rightarrow \infty}(-N^{-1}T \ln Z_G)=
\left\{
\begin{array}{lcl}
-T(\ln d-\frac{\varepsilon_g^2}{2\delta^2}) +\varepsilon_g &{\rm for} & T<-\frac{\delta^2}{\varepsilon_g},\\
-T\ln d -\frac{\delta^2}{2T}&{\rm for} & T\geq-\frac{\delta^2}{\varepsilon_g}.
\end{array}
\right.
\end{equation}

Let us compare the low-temperature behavior of the exact free energy $f$ and the approximation~$f_{G}$. First of all note that, at $T=0$, $f_G=\varepsilon_g$   as it should. One can easily verify that it is the choice of ground state energy as a lower limit of integration in eq. \eqref{Z from eta} which ensures that $f_G$ at zero temperature is always equal to the ground state energy per spin.

What about the entropy at zero temperature? GA gives $S_G\equiv\frac{\partial f_G}{\partial T}|_{T=0}=-\ln d+\frac{\varepsilon_g^2}{2\delta^2}$. However, according to  the third law of thermodynamics, entropy is zero at $T=0$ \cite{zem}.\footnote{The third low can be violated only for systems whose ground state degeneracy grows exponentially with the number of blocks.}
This disagreement shows how GA is unable to reproduce exact thermodynamics when temperature is low enough.\\
Note further that
\begin{itemize}
\item $f_G$ and $\frac{\partial f_G}{\partial T}$ are continuous at  $T=-\frac{\delta^2}{\varepsilon_g},$ however $\frac{\partial^2 f_G}{\partial T^2}$ and further derivatives are not. Thus GA predicts a fake thermal phase transition. This shows how, even at intermediate temperature, the agreement with exact results is very poor.
\item  $f_G$ at $T>-\frac{\delta^2}{\varepsilon_g}$ coincides with the first two terms of high-temperature expansion of $f$, as expected from \eqref{hightexp}.
\end{itemize}

\subsection{A generalized formula for the free energy under weaker constraints on Gaussian parameters}
Let us now relax one of the two constraints imposed on Gaussian parameters in the previous subsection and let us consider the situation in which $\Delta_{a}={\rm const}$ but $\overline{E_{a}}$ is not.

With the help of \eqref{sumint} we can rewrite  eq. \eqref{dens} in an integral form:
\begin{equation}\label{densintegr}
\eta_G(E)= \int \eta_{free}(\overline{E}) \frac{1}{\sqrt{2\pi N}\delta} e^{-(E-\overline{E})^2/(2\delta^2 N)}d\overline{E}.
\end{equation}
Here the "free" density of states $\eta_{free}(\overline{E}),$  introduced in \eqref{sumint}, is itself expressible as a summation $\sum_{\{ |a\rangle \}}\delta(\overline{E}-E_a)$. It is often useful to approximate it by some more regular and manageable function. In what follows we approximate it as a gaussian function (as often done when dealing with these kind of densities \cite{gauss}), exploiting a second time the Central Limit Theorem on which GA is based, thus introducing a further source of error since the gaussian is unable to reproduce exactly the behavior of $\eta_{free}(\overline{E})$ at its tails. We call the approximation thus obtained Double Gaussian Approximation (GGA), stressing the fact that the CLT is applied twice in two different steps of its derivation. Being aware of the further error thus introduced, let us develop the GGA. Let us assume that single-block free Hamiltonians $H_\mu$ are of equal form for all $\mu$ (generalization to the case of different $H_\mu$ is straightforward). Consider a basis $\{| a_\mu^l \rangle\},~l=1,2,..., d$ in a $d$-dimentional single-block Hilbert space. Without loss of generality we may put
\begin{equation}
\sum_l \langle a_\mu^l | H_\mu | a_\mu^l \rangle =0.
\end{equation}
The {\it statistical} standard deviation of a single-block self-Hamiltonian relative to a specific basis $\{| a_\mu^l \rangle\}$ is then defined as
\begin{equation}
\sigma^2\equiv \frac{1}{d}\sum_l \langle a_\mu^l | H_\mu | a_\mu^l \rangle^2.
\end{equation}
From the Central Limit Theorem we thus obtain
\begin{equation}\label{etafree}
\eta_{free}(\overline{E})\simeq \frac{d^N}{\sqrt{2\pi N}\sigma} e^{-\overline{E}^2/(2\sigma^2 N)}.
\end{equation}
Note that, as already pointed out, such an approximation is not a good one at the tails of the gaussian. However exploiting it, and within its limits, it is possible to apply the GGA to obtain
\begin{equation}\label{densityofstatesG2}\begin{split}
\eta_G(E) &\simeq \int_{\varepsilon_1}^{+\infty} d\overline{E}  \frac{d^N}{2\pi N \sigma \delta} e^{-(E-\overline{E})^2/(2\delta^2 N)-\overline{E}^2/(2\sigma^2 N)}\\
&=\frac{d^N e^{-\varepsilon^2 N/(2(\delta^2+\sigma^2))}}{2\sqrt{2\pi N(\delta^2+\sigma^2)}}\mathrm{erfc}\bigg[\sqrt{\frac{N}{2}}\sqrt{\delta^{-2}+\sigma^{-2}}\Big(\varepsilon_1-\frac{\varepsilon}{1+\delta^2/\sigma^2}\Big)\bigg].
\end{split}\end{equation}
Here $E\equiv N \varepsilon$ and $\varepsilon_1\equiv\min\limits_{\{| a_\mu^l \rangle\}}\langle a_\mu^l | H_\mu | a_\mu^l \rangle.$  Note that the case studied in the previous subsection, namely $\overline{E_a}={\rm const}$, corresponds to the limit $\sigma\rightarrow 0$. In this limit $\mathrm{erfc}(x)\sim 2$ and eq. \eqref{densityofstatesG2} actually reduces to the expression given in \eqref{densconst}, as it should.
In the framework of this approximation and using the asymptotic expressions for $\mathrm{erfc}$ one obtains
\begin{equation}\label{zG}\begin{split}
Z_G
&\simeq \theta\bigg(\varepsilon_1\Big(1+\frac{\delta^2}{\sigma^2}\Big)-\varepsilon_g\bigg) N \int_{\varepsilon_g}^{\varepsilon_1\big(1+\frac{\delta^2}{\sigma^2}\big)} \Bigg[\frac{d^N e^{-\frac{\varepsilon^2 N}{2(\delta^2+\sigma^2)}-\beta\varepsilon N}}{2\sqrt{2\pi N(\delta^2+\sigma^2)}} \times \\
&\frac{e^{-\frac{N}{2}(\delta^{-2}+\sigma^{-2})\big(\varepsilon_1-\frac{\varepsilon}{1+\frac{\delta^2}{\sigma^2}}\big)^2}}
{\sqrt{\frac{\pi N}{2}}\sqrt{\delta^{-2}+\sigma^{-2}}\Big(\varepsilon_1-\frac{\varepsilon}{1+\frac{\delta^2}{\sigma^2}}\Big)}\Bigg] d\varepsilon+N \int_{\varepsilon_{m}}^{+\infty}\frac{d^N e^{-\frac{\varepsilon^2 N}{(2(\delta^2+\sigma^2))}-\beta\varepsilon N}}{\sqrt{2\pi N(\delta^2+\sigma^2)}}d\varepsilon,
\end{split}\end{equation}
where $\theta(x)$ is the Heaviside step function and $\varepsilon_{m}\equiv \max
\Big\{
\varepsilon_g, \varepsilon_1\big( 1+\frac{\delta^2}{\sigma^2} \big)
\Big\}.
$\\
The first term corresponds to the $\mathrm{erfc}(x)$ with positive argument, while the second term corresponds to the $\mathrm{erfc}(x)$ with negative argument. There are some cases where the first term can be  neglected or equals zero (e.g. in Ising model). The second term alone gives rise to another complementary error function which closely resembles the one analyzed in the previous subsection in order to obtain eq.\eqref{fG}. Following the same steps we can then get
\begin{equation}\label{fG2}
f_{G}=
\left\{
\begin{array}{lcl}
-T(\ln d-\frac{\varepsilon_m^2}{2(\delta^2+\sigma^2)}) +\varepsilon_m &{\rm for} & T<-\frac{\delta^2+\sigma^2}{\varepsilon_m},\\
-T\ln d -\frac{\delta^2+\sigma^2}{2T}&{\rm for} & T\geq-\frac{\delta^2+\sigma^2}{\varepsilon_m}.
\end{array}
\right.
\end{equation}

Equation \eqref{fG2} shows the same behavior as \eqref{fG}. All comments made for \eqref{fG} are thus valid also in this case.

It is worth stressing that \eqref{fG2} is valid as long as the first term in \eqref{zG} can be neglected or actually equals zero, and within the limits of the additional approximation used to obtain \eqref{etafree}.

\section{\label{sec specific models}Application to specific models}
As an example of application of the Gaussian method to physical systems, we study its predictive power for the quantum Ising chain, extensively studied in literature \cite{lieb,mccoy}.
\subsection{Gaussian Approximation applied to the Ising model}
Quantum Ising model describes interacting qubits embedded in an external transverse magnetic field. It is probably the simplest quantum integrable model available. Its Hamiltonian reads
\begin{equation}\label{hamis}
H_{Is}=-J\sum_{j=1}^{N}\sigma_{j}^{x}\sigma_{j+1}^{x}+h\sum_{j=1}^{N}\sigma_{j}^{z},
\end{equation}
where $\sigma_{j}^{\alpha}$, $\alpha=x,y,z$ is the $\alpha$-component of the Pauli matrix vector corresponding to site $j$, $J$ is the spin-spin interaction parameter and $h$ is the external magnetic field. In what follows we mainly will work in the thermodynamic limit, where different boundary conditions yield the same result.

Being integrable, this model can be exactly diagonalized via a Jordan-Wigner transformation \cite{lieb}, obtaining the following expressions for ground state energy and for free energy per spin in the thermodynamic limit
\begin{equation}\label{Ising}\begin{split}
\varepsilon_g^{Is}&=-\frac1\pi \int\limits_0^{\pi} dp \sqrt{J^2+h^2+2hJ\cos p},\\
f^{Is}&=-\frac T\pi \int\limits_0^{\pi} dp \ln \left[2\cosh\frac{\sqrt{J^2+h^2+2hJ\cos p}}{T}\right].
\end{split}\end{equation}

Let us introduce the following notation for the eigenstates of $\sigma^{y}$
\begin{equation}
\sigma^y|y\rangle=|y\rangle,~~~\sigma^y|\tilde y\rangle=-|\tilde y\rangle,
\end{equation}
and analogously for $x$ and $z$.\\
In this notation $|z\rangle\equiv|\uparrow\rangle,~|\tilde z\rangle\equiv|\downarrow\rangle$.

To apply GA, we take as factorized states the eigenstates of $\sigma^y$, e.g.
\begin{equation}\label{a}
|a_y\rangle =|yy\tilde y\tilde yyy\tilde yyy...y\rangle,
\end{equation}
easily obtaining $\overline{E}_a=0$ and $\delta^2=J^2+h^2$.
We now may apply formula \eqref{fG} to obtain $f_G^{Is}$, which reads
\begin{equation}\label{fis}
f_G^{Is}=
\left\{
\begin{array}{lcl}
f_G^{Is-}&{\rm for} & T<-\frac{J^2+h^2}{\varepsilon_g^{Is}},\\
f_{G}^{Is+}&{\rm for} & T\geq-\frac{J^2+h^2}{\varepsilon_g^{Is}},
\end{array}
\right.
\end{equation}
where
\begin{equation}\begin{split}
f_G^{Is-}&=-T\ln2-\frac{T}{(J^{2}+h^{2})\pi^{2}}\bigg( \int_{0}^{\pi}dp\sqrt{J^{2}+h^{2}+2hJ\cos p} \bigg)^{2}-\frac{1}{\pi}\int_{0}^{\pi}dp\sqrt{J^{2}+h^{2}+2hJ\cos p},\\
f_{G}^{Is+}&=-T\ln2 -\frac{J^2+h^2}{2T}.
\end{split}\end{equation}
As expected, this equation predicts a wrong behavior at low temperature, as can be readily checked in Fig. \ref{isingfig}.\\

\subsection{Gaussian Approximation for the Ising model in a different factorized basis}
Let us now consider a different factorized basis, namely the eigenbasis of $H^0=h \sum_{j=1}^{N}\sigma_j^{z}$, which is the same basis used in \cite{lim} to numerically evaluate the partition function from equation \eqref{part}. With the same notation introduced in the previous subsection we set
\begin{equation}
|a_z\rangle =|zz\tilde z\tilde zzz\tilde z...z\rangle.
\end{equation}
This gives
\begin{equation}
\varepsilon_1=-h,~~~\delta_z=J,~~~\sigma_z=h
\end{equation}
and
\begin{equation}
\overline{E_{a}}=h(2k-N),
\end{equation}
where $k$ is the number of up spins in the factorized state $|a_z\rangle$ or, which is the same, is the number of $z$ in the state $|a_{z}\rangle$. $\overline{E_{a}}$ in this case actually depends on $a$. Equation \eqref{fG2} might thus be applied, provided its conditions are met.\\
One can check numerically that
\begin{equation}
\varepsilon_1(1+\delta_z^2/\sigma_z^2)-\varepsilon_g^{Is}\leq0
\end{equation}
for all values of model parameters $J$ and $h$. Thus free energy is given by eq.(\ref{fG2}). However, as can easily be checked, this result exactly coincides with one obtained in the previous subsection employing the y-basis.
The reason is that when we take a $y$-product basis as done before we get
\begin{equation}
\delta_y=\sqrt{J^2+h^2}
\end{equation}
and as a consequence
\begin{equation}
\delta_y^2=\delta_z^2+\sigma_z^2.
\end{equation}
This relation, here shown for the Ising chain, indeed holds for a certain class of systems as expressed by the following \\ \\
{\bf Statement:} Consider a model with Hamiltonian \eqref{block hamiltonian} represented as $H=H^0+I,$ $I$ being an inter-block interaction, and two product bases, $\{|a\rangle\}$ and $\{|\tilde a\rangle\}$, such that
\begin{enumerate}
\item $\Delta_a={\rm const}$ and $\Delta_{\tilde{a}}={\rm const},$
\item $\forall a~~ E_a=0,$
\item $\forall \tilde a~~ \langle \tilde a | I |\tilde a \rangle=0$ and $H^0|\tilde a\rangle=E_{\tilde a}^0|\tilde a\rangle.$
\end{enumerate}
Then
$$
\tilde\delta^2+\tilde\sigma^2=\delta^2.
$$

The above Statement implies that if conditions 1,2 and 3 are met for a particular system, then equation \eqref{fG} and equation \eqref{fG2} actually give the same result and one is entitled to use the simpler one. We might have stated this Statement right after equation \eqref{fG2} in the Section devoted to general results. However, for the sake of clarity and to better show its meaning, we decided to formulate such a Statement together with a concrete example. \\ \\

\begin{figure}[h]
\begin{center}
\includegraphics[width=250pt]{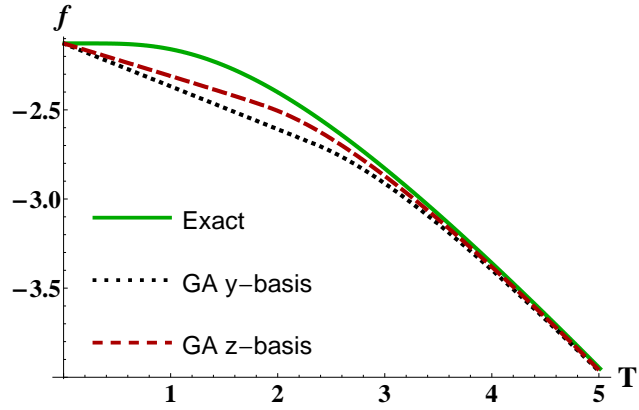}
\end{center}
\caption{Free energy per spin plotted against temperature obtained from the exact formula \eqref{Ising} (green solid line), from GA in y-basis (obtained both from \eqref{fis} or from the same expression stemming from \eqref{fG2}, black dotted line) and from GA in z-basis (obtained directly with the use of \eqref{part} in the case of $N=1000$, red dashed line) for an Ising chain when $J=1$ and $h=2$. It is possible to notice, as expected, how the high temperature behavior predicted by GA closely resembles the exact one, while at low temperature GA in both bases violates the third law of thermodynamics. Nevertheless the agreement with the exact result is better for GA in z-basis.}
\label{isingfig}
\end{figure}

We plot $f^{Is}$ and $f_G^{Is}$, obtained both from a GA in y-basis and from a GA in z-basis, on Fig. \ref{isingfig}.
One can notice a very good agreement between both the approximate free energies and the exact one as long as temperature is high enough ($T\gg\frac{-\delta^{2}}{\varepsilon_g}$), as expected from the high temperature expansion, equation \eqref{hightexp}, but such a good agreement breaks down at intermediate and low temperatures. GA in y-basis, in particular, gives the worst approximation, while GA in z-basis is closer to the exact free energy, while still showing all the drawbacks highlighted in the previous Section. Note that GA in y-basis, as shown, actually coincides with GGA for this particular model and thus is affected by a greater source of error, while GA in z-basis, as derived directly from \eqref{part} (and as such, not in the thermodynamic limit), is more accurate. The difference between the approximate and exact results is even more marked if we consider that all measurable quantities are obtained as derivatives of the free energy with respect to temperature. GA thus predicts, even in the thermodynamic limit, a behavior which is, at low temperature, very different from the one actually shown by Ising chain.

Note also that GGA can be applied in $y$-basis also (with $\delta_y=h$), however the dependence of $\overline{E_a}$ over $|a\rangle$ is more complicated in this case.
\subsection{Other models}
In the present subsection we give examples of other models to which the GA and GGA can easily be applied (showing, however, all the drawbacks discussed above).
\begin{itemize}
  \item cluster Ising chain \cite{clust}:\\
   $H_{Cl}=-J\sum_{j=1}^{N}\sigma_{j-1}^{x}\sigma_{j}^{z}\sigma_{j+1}^{x}+\lambda\sum_{j=1}^{N}\sigma_{j-1}^{y}\sigma_{j}^{y}$. In this case one can apply both GA according to eq. \eqref{fG} (in $x$- or $z$-basis)  and GGA according to eq. \eqref{fG2} (in $y$-basis), and easily check that the three conditions of the Statement actually hold. Thus GA and GGA coincide in this case. This fact, and further the fact that the application of GA yields exactly the same result as the one obtained for the Ising chain in the previous section are not surprising: it has indeed been shown that the Cluster Ising model is equivalent to an Ising chain in external magnetic field, and as such also GA can be applied equivalently.
  \item cluster Ising chain in magnetic field:\\ $H_{MCl}=-J\sum_{j=1}^{N}\sigma_{j-1}^{x}\sigma_{j}^{z}\sigma_{j+1}^{x}+\lambda\sum_{j=1}^{N}\sigma_{j-1}^{y}\sigma_{j}^{y}+h\sum_{j=1}^{N}\sigma_{j}^z$.
      In this case one can apply GA in $x$-basis and GGA in $y$- or $z$- basis. The conditions of the statement again hold if one properly divide the total Hamiltonian into the free part and the interaction.
  \item an alternate XY model:\\ $H_{AXY}=J_{x}\sum_{i=0}^{\frac{N}{2}-1}\sigma_{2i+1}^{x}\sigma_{2i+2}^{x}+J_{y}\sum_{i=1}^{\frac{N}{2}-1}\sigma_{2i}^{y}\sigma_{2i+1}^{y}.$ GA and eq. \eqref{fG} can be applied in the $z$-basis. Particularly interesting, this model has an highly degenerate ground state, which might lead it to violate the third law of thermodynamics and thus to a non-zero value for $\frac{\partial f_{AXY}}{\partial T}|_{T=0}$. However one in generals expects the zero-point entropy to depend only on the degree of degeneracy of the ground state, while the value predicted by GA depends on the energy of the ground state, which is clear from eq. \eqref{fG}. Even in this case we then expect the exact solution to be very different from GA in the limit of low temperature.
\end{itemize}
\section{Remarks and conclusions}
One could be wondering whether GA might be more efficiently exploited to get an insight on the physics of statistical models when analyzed in their non-thermal parameter space. The possibility of detecting signatures of quantum critical points (QCPs) in the behavior of thermodynamic functions is of great interest in the context of quantum phase transitions (QPTs) \cite{sac} and in connection with quantum correlations between sites. It is well known that an Ising chain described by \eqref{hamis} indeed shows the presence of a QCP when $h=J$, leading to a transition from a ferromagnetic to a paramagnetic phase. Such a transition is a second order QPT, which means it leaves signatures on the critical behavior of the second derivative of free energy with respect to one of the two Hamiltonian parameters when $T \rightarrow 0$. Such a criticality is due to a non-analytical point in the ground state energy (which by definition is the limit of free energy as $T\rightarrow 0$) and is thus clearly present, as easily checked, also in the GA which strongly depends on $\varepsilon_{g}$. Nevertheless, as shown in the previous section and as can be seen from \eqref{fG}, the knowledge of $\varepsilon_{g}$ is a necessary condition for the application of GA to any system. This means that in order to calculate $f_{G}$ one should know in advance $\varepsilon_{g}$ and thus its whole set of non-analytical points. GA is then unable to supply further information on QPTs or to extract them more efficiently directly from the Hamiltonian operator. Differently from what argued in \cite{lim} then, GA can actually be applied to the study of QPTs (indeed at $T=0$ the free energy, as given by GA, coincides with the true one even if, as noticed in \cite{lim}, the two partition functions actually differ) but requires the same computational effort as an exact analysis and is thus not useful in simplifying any calculation.\\

The prime goal of our study has been the investigation on the range of validity of the GA. The larger is the system, the better GA should work, therefore we mainly concentrated on the thermodynamic limit. We have shown that in this limit GA leads to simple expressions for thermodynamic functions in two important cases, namely when the temperature is high enough and when a product-state basis $\{ |a\rangle \}$ exists for which the standard deviation of the Hamiltonian $\Delta_a$ does not depend on $|a\rangle$. Our results can be summarized as follows:
\begin{itemize}
  \item At high temperature any thermodynamic quantity as given by GA coincides with the high-temperature expansion of the exact function, up to terms of the order of $\beta^2$.
  \item For systems with $\Delta_a$ independent on $|a\rangle$ and for systems with finitely-many blocks GA predicts nonzero entropy at $T=0$, in contradiction to the third law of thermodynamics.
  \item For systems with $\Delta_a$ independent on $|a\rangle$ GA predicts a fake thermal phase transition at an intermediate temperature, in contradiction to the general result that thermal phase transitions are absent in 1D systems with finite-range interactions \cite{1dsystems}.
  \item As discussed above in this section, GA is useless when applied to the analysis of quantum criticality at a fixed temperature.
\end{itemize}
In conclusion, we have provided convincing arguments showing that GA, generally speaking, fails at low and intermediate temperature, and is thus a pretty weak approximation tool.

\begin{acknowledgements}
OL acknowledges the partial support from grants NSh-4172.2010.2, RFBR-11-02-00778, RFBR-10-02-01398 and from the Ministry of Education and Science of the Russian Federation under contracts N$^{\underline{\rm o}}$ 02.740.11.0239.
\end{acknowledgements}

% BibTeX users please use one of
%\bibliographystyle{spbasic}      % basic style, author-year citations
%\bibliographystyle{spmpsci}      % mathematics and physical sciences
%\bibliographystyle{spphys}       % APS-like style for physics
%\bibliography{}   % name your BibTeX data base

% Non-BibTeX users please use

\end{document}